# Teaching Physics to Life Sciences Students in the SCALE-UP Style

Ganga P. Sharma,[1] Shantanu Chakraborty,[2] and Bilas Paul[3]

[1] *Fairmont State University Fairmont, WV 26554, USA*

[2] *Valdosta State University, Valdosta, GA, 31698, , USA*

[3] *SUNY Farmingdale State College, Farmingdale, NY 11735, USA*

Physics has a reputation among majority of life sciences students for being very complicated and tough. If we leave students with this impression, it is likely that students see physics class as useless and irrelevant to life sciences. Concepts of physics are vital in oder to understand physics based technological tools and biophysical topics essential and relevant for life sciences. This review summarizes approaches for improving teaching and learning in introductory physics courses to life science students in the SCALE-UP style. We also discuss our experiences in adapting IPLS Courses to better meet the needs of life sciences students.

## I. INTRODUCTION

In recent decades, leading researchers in life and health sciences have been pointing out importance regarding reformation of introductory physics for life sciences (IPLS) courses that are essential to improve scientific skills and competencies among scientists and professionals in their fields. The aim of life sciences is to learn everything about life on this planet. Therefore, the life sciences are so complex with many fields such as biochemistry, botany, cell biology, cognitive neuroscience, ecology, genetics, health sciences, medicine, microbiology, zoology, and a lot more. It is apparent that physics and life sciences see the world differently, but we cannot separate them. As physics deals with energy, matter, space and time which are also very important components in life sciences because they are also basis for living organisms. It is realized that understanding of biological matters, their functions, and interactions are impossible if they are not completely understood on the level of physics.

In early 1950s, American biologist James Watson and British physicist Francis Crick proposed the double helix structure of the DNA molecule. With their famous model of the DNA double helix, they marked a milestone in the history of science and gave birth of modern molecular biology. This was the first time the world witnessed power of collaborative and interdisciplinary research, especially when researchers with expertise on life sciences and physics work together. Watson and Crick mostly collected and analyzed existing pieces of X-ray crystallography data taken by Rosalind Franklin to predict correct structure of the DNA molecule. A father and son team, both called William Bragg, invented X-ray crystallography in 1912 and Nobel prize in physics was awarded to them for this discovery in 1915. X-ray crystallography now has developed into an indispensable tool for structural biologists. This is one of the examples on how physics can make important contributions to life sciences. We cannot imagine modern biology and medicine without physics-based technological tools, such as stethoscope, ultrasound, acoustic microscope, nuclear magnetic resonance imaging (MRI), positron emission tomography (PET), laser, linear accelerator and many more. Scientists are trying to develop a unified understanding of how genes are regulated, and researchers are attempting to understand the dynamics and functional activity of the brain by applying ideas from physics.

Knowledge of fluid mechanics, the branch of physics concerned with the mechanics and forces on liquids and gases, is essential to better understand about human cardiovascular system, including blood pressure, blood flow and resistance. In the same way, we can take countless examples that can correlate laws of physics to understand the rules of living systems. For this reason, about 93% of US medical schools and other health-related professional programs require or recommend successful completion of college-level introductory physics courses (equivalent to 8 credits) to meet the admissions requirement [AAMC 2014]. The successful completion of introductory physics for life sciences (IPLS) courses are required when it comes to enroll in graduate schools for a biology degree as well. Therefore, IPLS courses are instrumental to those students who intend to pursue basic research careers in the life sciences as well as those who aim for professional education in medicine and other health-related fields.

The following sections focus on various aspects of promoting active learning in educational settings. They cover Student Centered Active Learning Environment with Upside-down (SCALE-UP) pedagogies (Sec. II), Active Learning Classrooms (Sec. III), Strategies to Motivate Life Sciences Students to Learn Physics (Sec. IV), and Active Learning Techniques (Sec. V). These sections aim to provide insights and approaches for creating engaging and interactive learning experiences that encourage student participation and deepen their understanding of the subject matter. This review mainly focuses on improving teaching for introductory physics for life sciences students but strategies explained here can be used for effective teaching for any of undergraduate courses.

## II. STUDENT CENTERED ACTIVE LEARNING ENVIRONMENT WITH UPSIDE-DOWN (SCALE-UP) PEDAGOGIES

Students majoring life sciences usually tend to struggle with basic mathematics or come with a background with deficiency in mathematics. For that reason, most students are not comfortable in physics. In fact, one cannot ignore mathematical foundations as a part of IPLS courses curricula. This creates a huge challenge for IPLS instructors in designing IPLS courses and classroom pedagogy that can be successful to develop students' quantitative and problem-solving skills with strong physical pictures relevant to biological context.

Traditional lecture method of teaching has been proven ineffective in terms of improvements in student performance. But sciences, including physics, can be learned effectively through active inquiry and exploration. Therefore, IPLS courses must be structured so that research based active-engagement pedagogies can be implemented in the classrooms to elucidate the complex situations found in living systems. In recent decades, many steps toward IPLS courses reform have been taken. Conferences, sessions and workshops focusing on IPLS courses have been held at the Summer and Winter meetings of the American Association of Physics Teachers (AAPT). These undertakings were influential in furthering IPLS reform at many institutions [1].

Students can most effectively learn a course when they actively engage in learning activities. Faculty can increase the level of student engagement by creating and implementing learning activities that can assure students about relevant connections between their current learning and their future. Research has shown that if students do not consider a learning activity worthy of their time and effort, they might not engage in a satisfactory way, or may even disengage entirely in response [2]. One of the goals in teaching IPLS courses is to help students build connections between their life science studies and physics which is not easy task because of tremendous diversity within the life sciences and the allied health fields. This diversity poses a complication in choosing appropriate topics and applications while designing instructional activities to accommodate needs of all IPLS students. In addition to this, another crucial thing is to make decisions about teaching pedagogies that make learning much more effective and student centric. SCALE-UP pedagogies significantly restructure the classroom environment and pedagogies to promote highly active and interactive instruction [3].

Inquiry-based learning in a studio-like technology-rich setting is the most salient feature of SCALE-UP model of instruction. The successful implementation of this learning environment over 500 colleges and universities across the US and around the world, including Massachusetts Institute of Technology (MIT), has shown significant results in students' learning outcomes and retention, including improvements in attendance rates and grades as well as enhancements in problem-solving skills and overall understanding. SCALE-UP was originally developed by Robert Beichner, Professor of Physics at North Carolina State University with aim to develop a model of teaching that could improve students' conceptual understanding more effectively than a traditional model of separate lectures and seminars/labs [4]. SCALE-UP has been instrumental in transforming large-enrollment introductory physics sequence for life-science students to a Lecture/Studio format and aligned the physics concepts with authentic biological applications [5]. Moreover, SCALE-UP has now been used to a variety of disciplines of various class sizes as a successful instructional model. Rooms with round learning stations are needed to facilitate interactions between small groups working on short, interesting tasks: hands-on activities, questions, simulations, or laboratories, all at the same location for this instructional environment. Two semester sequence of IPLS courses are needed to introduce life science majors to the concepts of physics in a biological context. Each course should be equivalent to three-credit lecture and one-credit lab combine for six hours of in-class work each week. Lecture and lab should be combined into a single session to allow more flexibility with time and activities. Students receive the same grade for both lecture and lab because of the interwoven nature of these two parts of the class. The class size for SCALEUP pedagogy typically ranges from about 27 to 117 students. Two instructors and teaching assistants may be needed to teach single section of IPLS courses depending on the class size. One of the instructors can act as a primary instructor who is also the instructor of record and ultimately responsible for planning the class, providing instruction and grading. As class-wide discussions are at the heart of the lectures, secondary instructor and a teaching assistant help to facilitate interactions between teams of students and engage and involve students as active participants. Having multiple instructors is a proven strategy to improve student attention and retention in physical classrooms at universities.

## III. ACTIVE LEARNING CLASSROOMS

Active Learning Classrooms (ALCs) accommodate teaching and learning activities by promoting active, collaborative and supportive atmosphere. Following principles need to be considered when designing or renovating classroom spaces [6].

- Learning space should create a conducive learning environment where students can engage with course materials and a range of classroom technologies to support multiple modes of learning. • ALCs should provide features that allow students to learn cooperatively and collaboratively. • Flexible learning spaces should enhance communication and interaction between students and faculty.

"Student-centered active learning spaces and classrooms feature round or curved tables with moveable seating that allow students interact with others. The tables are often paired with their own whiteboards for brainstorming and diagramming" [7]. Hence interactive projectors, displays, good tables, moveable chairs, and lots of whiteboards are essential ingredients to support active learning environment. Collaboration and content sharing technologies in the classrooms also play vital roles for student success.

## IV. STRATEGIES TO MOTIVATE LIFE SCIENCES STUDENTS TO LEARN PHYSICS?

### A. Increase Relevance

The development of interesting instructional content and activities that increase relevance and usefulness to life sciences applications is of the utmost importance in motivating IPLS students. The decisions taken by an instructor to deliver the content using most appropriate method or a combination of methods should be aligned with course goals, assessment practices and criteria, and the nature and backgrounds of the students. The IPLS students may not have physics background.

The lack of students' motivation in learning physics is not surprising because only few of IPLS students have significant previous experience with physics. An instructor can motivate IPLS students by designing and delivering instructional activities that inform basic physics concepts with application to the life sciences. We can unlock IPLS students' internal drive for learning when they are being taught in a manner in which students see how physics plays a significant role to understand biological systems and matters. Giving students the opportunity to solve life sciences problems using concepts and ideas of physics is the ultimate source of motivation. It is, therefore, imperative to create an environment where students must fell both interested and excited about how and why they are learning physics. Reformation of the IPLS courses has been ensured by integrating physics in the life sciences contexts. IPLS instructors should select most important physics content in biological context to prepare and motivate life science students. Let us take some examples that make interdisciplinary connections between introductory physics topics with life sciences. • Kinematics: cell movement in cancer invasion and metastasis, diffusion in cells • Forces: cell signaling, cell motility, DNA replication and repair, cell growth/division

- Torque: muscle functions
- Fluids statics: blood flow and resistance, barometer, lung pressure, buoyancy in real life, hydrostatic weighing, erythrocyte sedimentation rate (ESR) testing
- Elasticity: aging, elastic proteins, orthopaedics • Waves: how hearing works, shock wave as biological therapeutic tools, bio-electromagnetic (BEM) devices, stethoscope, ultrasound, echolocation
- Light and Optics: light microscopy, laser therapy, fiber optics
- Electricity: nerve signal transport, bioelecticity, insertion of drugs into tumors and DNA into cell nuclei using electric fields, cell membrane potential • Magnetism: nuclear magnetic resonance imaging, magnetic sensing
- Heat and Thermodynamics: thermoregulation of animals, metabolism

Individual IPLS faculty should improve their courses by reviewing and modifying courses with addition of authenticate life sciences problems. Increasing relevance in IPLS courses requires great insight and passions in physics and life sciences. Relevance theory of cognitive science [8] explains basic features of human cognition. According to this theory search for relevance is one of the essential components to motivation. Making content relevant to students' personal and career goals not only motives students, but also makes learning process fun and engaging [9].

### B. Concept Test as an Early Intervention

Even though mathematics, including trigonometry and algebra, are prerequisites required for IPLS courses, significant numbers of students may lack the appropriate mathematical background to cope with introductory physics courses. Students who are weak in math have low self-confidence in their physics work and are intimidated by physics equations. Deficiency in quantitative skills is associated with higher failure rates in these courses, even if students possess good qualitative reasoning skills. Those IPLS students who perceived mathematics as a difficult subject tend to leave disciplines. This leads to low student retention and eventually, results skilled employees shortage that will threaten life sciences industries, including fundamental life sciences research, healthcare, and biopharma.

Many life sciences students taking physics courses enter with the firm belief that they will achieve excellent or superior grades in physics courses. Which is not surprising because securing better grade helps these students to pursue and advance careers in the biological or health related sciences. There have been numerous studies which show a positive correlation between students' mathematical skills and their exam grades in physics courses. However, poor mathematical skills and lacking understanding the problem are more prevalent among IPLS students. Therefore, IPLS instructors should come with a plan that can identify the students who are weak in basic mathematics early in the semester. On the other hand, instructor should provide extra assistance that enables weak students to excel in mathematics. Inviting students during office hours for the resolution of exercises and concept enforcement is the best option for this purpose. Providing information about available free resources, such as tutoring services, and encouraging repeatedly working on problems help students to succeed.

Concept tests that are administered during beginning of the semester are proven to be effective to assess students' prior knowledge. Numerous studies have shown that student's prior knowledge directly impacts their learning in class (Nagy, Anderson, & Herman, 1987; Bloom, 1976; Dochy, Segers, & Buehl, 1999). By systematically using concept tests to accurately gauge quantitative skills of IPLS students at the beginning of semester, educators can address problems early on through various interventions and identify individual needs, promotes deeper understanding, and ultimately helps students overcome their mathematical deficiencies. Authors experiences on concept tests and early interventions also show that students who got extra help after concept test did better than those who didn't receive extra support as a form of early intervention. Every detail of this study is not part of this review, but data is being analyzed and we expect to publish in near future.

### C. Diversity and Inclusive Teaching

The foremost thing to incorporate diversity in the classroom means recognizing that all students in the classroom are unique in their own way in terms of their abilities, skill sets and talents. The instructor should have this understanding to develop varied course materials, teaching methods, and learning activities that accommodate a diverse group of students with a range of learning styles, abilities and experiences.

What happens in the classroom plays a significant role to enrich the learning experience. Every faculty member can take actions to make a difference in the classroom by creating an inclusive, highly effective and conducive learning environment. The primary aim of such an environment is to establish a greater sense of belongingness in class. That sense assures that all students in class fell welcomed, affirmed, respected, and valued. Faculty must show respect for all comments and questions from students and provide an atmosphere where students feel safe to contribute. Striving to connect with and reach out to struggling students is the best strategy to keep students connected and engaged in learning process. It is imperative to explain context of references and examples before using them to make sure that all students in the classroom are on the same page and no students are left behind. This is very important because everyone in the classroom may not be familiar with the context of that references/examples.

Incorporation of Universal Design for Learning (UDL) principles gives all students an equal opportunity to learn. For example, presenting materials both orally and visually can accommodate both auditory and visual learners and students with sight and hearing disabilities. Multiple means of instructional content in the form of text, audio, video as well as hands-on activities provide flexible and accessible opportunities for all students that best suit to their learning strength. Students with disabilities may struggle with large class sizes, fast pace of instruction, lack of scaffolding in the curriculum, precision of the content, and the pedagogical approach from faculty. Universal Design for Learning has been playing crucial role in overcoming these barriers [10].

Students from underrepresented groups can be too shy to speak up and interact with other fellow students and with professor. They may be scared of embarrassment and being put on the spot. They may feel like someone will make fun of them later. Such stereotype feeling directly demotivates learners and impacts their learning and academic performance [11]. Helping students to overcome these emotions is important to ensure students become more comfortable with their peers for effective classroom discussions. Motivation is one of the main foundations of an effective classroom. This can be achieved by providing right instructional resources and implementing right instructional methodologies at the right time, in the right medium, and at the right pace.

As IPLS students are one of the most diverse student populations in terms of their major course of study, it is valuable to provide a variety of perspectives on each topic of instruction. The simple gesture of addressing a student by their name breaks the ice for the college classroom. So, investing time to focus on getting know about students contributes to an inclusive learning environment [12]. Increasing numbers of reports and articles have been published in recent years showing that integration of diverse and inclusive teaching not only motivates students, but also promotes critical thinking, problem solving, and creativity.

### D. Effective Educational Videos

Effective educational videos have long been used as an educational tool for motivated and successful learning and are integral part of higher education. Several studies have indicated that educational videos serve as a productive part of a learning experience [13–15].

John Sweller proposed the very first Cognitive Load Theory (CLT) in 1988 which explains how human mind processes information. According to this theory, neurons form long-term memories by processing auditory and visual mode of information more efficiently than other everyday observations. For example, electricity is very important for the nervous system to transmit signals throughout the body and to the brain. Electric fields are an important tool in understanding how electricity begins and continue to flow. Hence electricity is one of the topics that is of particular importance to the life sciences students. The concept of electric fields and their properties, for instance, is easier to comprehend when demonstrated visually using a video. Including a video with Interactive simulation aids IPLS student comprehension on such a physics concept. Research studies recommend short video lessons (about 5 to 10 minutes) that are proven most effective to support student learning [16, 17].

Educational videos are important content-delivery tool when it comes to flipped, blended, and online classes. A short video about a specific topic is prepared and is made available to students via a learning management system, such as Moodle. Students watch the video and absorb the concepts on their own time before class meets. Then class time is spent doing activities, solving problems and working in small groups to enhance joint intellectual effort. When students work together in groups, they bring multiple perspectives in solving problems. This approach of teaching not only enables an instructor to evaluate immediately how students are learning, but also allows to provide immediate feedback on their progress. Fostering student motivation is the key in all these activities. It is worthy to note that only motivated students tend to engage in learning activities, which in turn lead to student success.

Faculty may assign educational videos to watch either in-class or outside of class depending on the instructional need. IPLS Instructor must use pictures, illustrations and animations more than mathematical expressions to explain any physics concept. Learning outcomes of the course dictate the best use of these videos. Planning, recording, and editing a high-quality educational video is time consuming, but it's worth making in terms of student comprehension. In addition to making videos shorter, Instructors should consider couple of other important aspects while designing effective educational videos that increase engagement (Guo, et al, 2014). Video must begin with clear introduction about what it will cover, and narration should be in conversational styles like in the storytelling. Studies have shown that speaking in formal style in the video does not serve better in terms of student engagement during learning process. Maintaining appropriate eye contact is also important while speaking to the camera just like in the physical classrooms. Because this is the only way of non-verbal communication which helps to build relationship between instructor and students. Good relationship between students and teacher has long lasting positive implications on students' academic performance.

### E. Choices

Learning activity that gives students a greater degree of control over their learning fosters engagement. Giving meaningful choices during learning process can help to bring out students' best effort and unlock their full potential. Making students in charge of their learning boosts motivation by meeting each student's individual needs as they have different interests and strengths. Offering student choices not only within assignments, but also in choosing learning materials and resources, such as textbook selection, create motivated learning culture that gravitates students toward positive learning experience. We can design multiple learning activities to explain same topic but in different context and allow students to choose one of these activities which best suits to their interest and enthusiasm. For example, consider two learning activities are designed related to Bernoulli's principle:

1. An activity that helps to understand how Bernoulli's principle is linked to cardiovascular physiology.

2. An activity that helps to understand how marine mammals achieve lift by exploiting Bernoulli's principle in nature.

If IPLS students have opportunity to choose one of these two activities, we can easily guess that first activity will likely to be chosen by IPLS students with Biology major whereas students with Marine Science major tend to choose second activity. Students become acquainted with concept and application of Bernoulli's equation in the real world through either of two activities. But when IPLS students with Biology major choose first activity and students with marine science major choose second activity, they are more joyfully engaged in the activity. It has been found that opportunities having choices motivate students by building ownership in the learning and motivated students process their understanding and store it in long-term memory more effectively [18].

Encouraging students ask questions and work through problems on their own incorporate students' choice in the classroom. Gathering feedback through a survey or other means to assess learners' acceptance and satisfaction with instructional materials, methods, activities and grading scheme, etc. is helpful to design and implement instruction as per students' choice. All these practices increase learners' autonomy which eventually transforms students into life-long learners.

## V. ACTIVE LEARNING TECHNIQUES

The use of an active learning approach in a SCALEUP style involves students in thinking, discussing, and investigating about problems and topics in the course materials [19]. One of the main characteristics of upside-down pedagogy is that it flips the traditional method of instruction which emphasizes recitation and memorization techniques to educate students. But in upside-down pedagogy, a student becomes a teacher at first by reviewing the day's basic materials before ever get to class. This includes reading, watching, listening, and noting course materials, taking quizzes and even working with simple problems before class. Secondly, class time is fully utilized working in more advanced applications and problems related to the topic. Interesting activities are assigned to students to investigate, and carefully structured groups are assembled to facilitate discussion. While students work in groups, the instructor is free to roam around the classroom–asking questions, sending one team to help another, or asking why someone else got a different answer [20].

Active learning is based on constructivist approach of learning. Constructivist approach of learning engages students actively to build their own knowledge and understanding of the world by combining new ideas with experiences and existing knowledge [21]. This has been demonstrated in a number of scientific studies that active learning approach is successful and effective in improving learning outcomes by promoting collaboration and engagement. Enhancement in critical thinking, problem solving and decision-making skills are major benefits of active learning approach. Multiple active learning techniques may be used based on instructional needs. Here are some major techniques which are highly effective in teaching IPLS students.

### A. Brainstorming

Brainstorming is one of the student-centered active learning techniques which is used to bring out a number of ideas from students to a specific topic/problem. In this technique, instructor invites students to think spontaneously and creatively by posting a question. Since the question is new to students, it challenges their current level of understanding on the problem. Then students are placed into groups to produce possible ideas and ways of solution to the problem. Discussion takes immediately after brainstorming session ends to find appropriate idea(s) for possible use as a solution toward solving the problem.

During brainstorming session, instructor acts as a facilitator who ensures full participation from all students. All ideas are welcome without criticism and evaluation to maximize students' contributions. Each group should consist of students with varied experiences, backgrounds, knowledge and academic disciplines to bring unique ideas for the discussion. Creating relaxed informal learning environment during the session, where students can collaborate rather than compete with one another, can lead to meaningful idea generation and idea building to solve the problem.

Instructor should move around and help students when they are stuck and not able to generates ideas. At the end of each session, thanking and appreciating students for their contributions encourage them to participate in a future activity. There are a number of ways that instructor can implement this technique of instruction in the classroom to improve its effectiveness [22]. The bottom line is any of these ways should stimulate students' interest in a subject by allowing free flow of ideas to develop problem-solving and higher-order thinking skills.

Brainstorming is great for idea generation in both online and face-to-face classes, during both synchronous and asynchronous schedules. However, there are some drawbacks of this method of instruction. This may take too much time if the group is not properly assembled and controlled. For example, some students in a group are shy to speak up and some speak too much. It requires ground rules to ensure that every single student's ideas are heard within stipulated time. 5 to 10 minutes are sufficient for the best ideas generation in majority of class activities. In some instances brainstorming may reduce the originality as some wrong ideas may also be accepted for discussion. Students can leave the session thinking that was waste of time. This feeling of the time being wasted can be reflected when students are not actively involved in future activities. But advantages of effective implementation of brainstorming session outweigh the disadvantages.

### B. Think-Pair-Share

The think-pair-share (TPS) is a cooperative learning technique that encourages individual learner think alone and provides an opportunity to discuss their thought with pair and share with groups and the entire class. This technique was first proposed by Frank Lyman of the University of Maryland in 1982. There are following distinct steps involved in TPS which include the following:

1. Firstly, the instructor presents an open-ended question for students to answer.

2. Secondly, students think independently about the question to generate ideas on their own.

3. Thirdly, students turn to the person next to them to form pairs and discuss their ideas or responses.

4. Lastly, student pairs share their ideas with a larger group, such as the whole class.

For TPS questions to serve their purpose, they must be conscientiously aligned with instructional goals. Therefore, special considerations need to be given during designing the questions for TPS. These questions must provide students the opportunity to learn and practice the knowledge and acquire skills, including thinking, listening, writing, communicating and collaborating, required for successful learning. This technique of instruction is quite simple in terms of preparation but surprisingly successful in terms of students' engagement and learning outcomes across all learning environments [23].

There are many IPLS topics students can discuss as TPS activity. For instance, if some students are struggling to understand about sound waves, initiating a discussion on how hearing works will be helpful to enhance students' comprehension on it. Discussion about structures of the inner ears and how they are related to sound perception is also a great topic. Another great topic would be what is ultrasound and how it is useful for medical applications? As students collaborate to reflect and discuss in a think-pair-share via such activities, higher-level understanding will be achieved.

Despite the advantages, the TPS has also some pitfalls. There is a possibility that one of the students in a pair totally dominates the conversation and other partner who has low understanding of the topic doesn't get an opportunity to share their thinking. This has a bad impacts on students' proactive, energetic, and sustained participation. It is therefore imperative to provide a clear instruction on how interactions take place and who talks first during TPS session. One of the pitfalls in this process is we can experience a lot of downtime in the classroom if a lesson is not fully prepared. Think-pair-share can be too noisy. A think-pair-share can take 3 to 10 minutes depending on question, activity and class size. The instructor should be aware of a fact that it may put time pressure on some students. Pressure has been shown to increase stress and anxiety in students. Therefore, the instructor should be flexible if students need more time to accomplish the task.

### C. Demonstration

Demonstration is a teaching method used to communicate an idea and exhibit the application of a concept using various activities, including classroom experiment, survey, simulations, poster presentation, and analysis of the secondary data, etc. Demonstration connects theories to actual applications which helps students to find relevance of their learning. Direct students' involvement is required throughout the entire process. After demonstration, students must present reflection about their learning and analysis of the results [24].

Demonstrations are valuable in the acquisition of conceptual skills in the event when students have a hard time to understand. This approach of teaching actually ensures each student obtains mastery in a given topic. Once Benjamin Franklin stated, "Tell me and I forget. Teach me and I remember. Involve me and I learn." These adages speak conspicuously about learning benefits of demonstration.

We can summarize steps involved in demonstrations as follows:

1. Firstly, the instructor describes the problem or shows the demonstration.

2. Secondly, students make prediction about the result of a demonstration and discuss with neighbor.

3. Thirdly, Students may conduct the demonstration in small groups for hands-on experience.

4. Lastly, students record data, analyze data and present the results.

There are a lot of factors that determine effectiveness of this method of instruction. It is important to make sure that each student predicts the result based on their understanding, even if it is incorrect. In this stage of the process, students should get an opportunity to share and discuss prediction they made with their partners. Integration of critical reflection after the demonstration is very important because it helps students to think explicitly about what they have learned. It is important to provide explanation on why prediction may have differed with the observed result.

Interactive simulations are highly effective for IPLS students to learn complex physics concepts in an entertaining way across all learning environments. There are so many ready-to-run simulations available freely in the internet that cover a wide range of introductory physics topics, such as Newton's laws of motion, conservation of momentum, collision, buoyancy, gravity, friction, electric field, waves and sound, heat transfer, etc… These simulations provide excellent in-class demonstrations and can be also easily customized to meet specific instructional needs.

Disadvantages of method of demonstration include shortage of time in terms of preparation during hectic teaching schedules. This method cannot be used in all types of teaching situations. It requires appropriate demonstration spaces and expensive resources such as equipment, software, etc.. There is always a greater risk of students being inactive throughout the process as instructor plays the main role in the demonstration. This method of instruction neglects individual differences among students. To resolve this problem, instructor can break the demonstration into smaller steps and involve students in each step, either by having them perform tasks or by asking for their input on the next step. Students should be allowed to explain parts of the demonstration to each other to foster deeper understanding and collaborative learning.

### D. Group Work

Engaging students to work together in groups is a very effective way to encourage them practice the skills they are learning. This technique is highly successful in inquiry-based learning. In this method, students work in a group to investigate worthy questions and tackle more complex problems than they could on their own. Working in groups not only develops critical-thinking, communication, and decision-making skills but also instrumental to acquire teamwork skills. Teamwork skills are not only important during learning but also essential to the workplace in this competitive world to increase productivity. Moreover, working in groups has been shown to contribute to student learning, retention and overall college success (Astin, 1997; Tinto, 1998; National Survey of Student Engagement, 2006).

There are a multitude of ways that the instructor can implement group works in the classroom. Each way must start with careful design of the group activity that can closely relate to the course objectives and class content to enhance students learn [25]. Then students should be arranged in groups, and it is necessary to make sure that all students feel comfortable with their group. As SCALE-UP classroom layouts are effective to facilitate discussions, there is no doubt students will be able to hear one another clearly for robust interactions. Each member of the group should be given designated roles, such as leader, prober, recorder, devil's advocate, presenter, etc., to hold them accountable.

Clear instruction should be given to students about what exactly they must do before they embark on a discussion. As a facilitator, the instructors' role when students do their work is to monitor their progress as well as answer the questions they may have. But there will be no interference from instructor's side with group functioning. This approach allows students to struggle with problem and helps to accomplish the task. Group activities should be concluded with reporting from each group to promote exchange of ideas.

This method of instruction inherits some disadvantages like other methods of instruction which need to be taken into consideration. If this strategy implemented incorrectly, it could lead to a situation where only 1-2 students taking the bulk of the workload, and other members may not contribute significantly by involving in other tasks or gossip. Therefore, the instructor should encourage students to define roles and responsibilities of each member of a group. If students within a group unable to communicate effectively, it can create conflicts among group members. By creating an inclusive and diverse learning environment, we can resolve this problem. The beautiful sense of diversity in the classroom would not only result effective learning environment but it would transform our society into a more prosperous one by developing diverse work force. Despite these limitations, each member of a group gets an opportunity to learn new things from each other, by paving the way to refine their understanding.

Group work can be time-consuming and may lead to inefficiencies if not managed well. Therefore, the instructor must provide clear timelines and milestones for group tasks to keep students on track. Students should be encouraged to break down tasks into manageable steps and assign deadlines for each step. Therefore, teaching students about time management techniques is essential to help them work more efficiently. Differences in communication styles and language proficiency can hinder effective collaboration. Hence educators must be mindful of forming diverse groups to ensure a balance of communication styles and to foster inclusivity.

## VI. CURRICULUM DESIGN

We discussed some of the active learning strategies in our previous sections. Now we focus on effective curriculum design process in the active learning environments that we implement in our teaching practice. Traditional instructor-centric method of instruction begins with designing different course materials and various class activities at first. Then tests, assessments and evaluation are used to measure how well student are learning the materials. This method focuses on what the instructor wants to do. Excessive lecturing and assessments without clear learning goals can take students away from the essence of the class. But student-centered Active Learning Environment with Upside-down Pedagogies (SCALE-UP) offer instructors opportunities to design curriculum backwards from the learning outcomes. Writers and curriculum experts Grant Wiggins and Jay McTighe developed a well-structured backward design of instruction [26]. The primary purpose of backward design is to help instructors and students for better planning and understanding of both the teaching and learning processes in transparent, efficient and engaged ways. There are three sequential stages of backward design. They are:

1. **Identifying Learning Outcomes**
   The first stage of backward design begins with identifying observable, measurable and reachable learning outcomes of a program, course or single lesson. This stage helps instructors to determine what contents and learning activities are necessary for students to achieve those outcomes. When students know learning outcomes of a course or lesson, it is more likely that they are in a better position for their learning quest. It is utmost important to incorporate three main domains of learning. These domains of learning are cognitive (thinking), the affective (social/emotional/feeling), and the psychomotor (physical/kinesthetic) domain. Learning outcomes should explicitly state expected knowledge, values, skills and attitude from students at the end of a course or lesson. It is important to use simple language with action verbs such as calculate, determine, solve, draw, interpret, apply, analyze, etc. to describe learning outcomes. The instructor must avoid verbs such as understand, learn, know, etc. because they are complex to measure or observe. The faculty may find difficulties in determining learning goals despite their expertise in the subject matter. Therefore, faculty must give considerable time to think about learning outcomes. Poorly written learning outcomes pose significant challenges in the assessment process. If learning outcomes consist of compound statements, it is prudent to include a short description that can explain how assessment procedure will identify student success. Before preparing a list of learning outcomes, following points are also essential to consider: • Learning outcomes should align with the program's curriculum.

- Learning outcomes should focus on learning products and not the learning process.
- There should be enough learning outcomes.

2. **Determine Acceptable Evidence**

Once the desired results are established, the next step is to design assessments that will provide evidence of student learning. In this stage, teachers and curriculum planners identify and develop evaluation and assessment methods/strategies to know whether students achieved the desired results. The compelling question at this stage is: what we will count as acceptable evidence of student understanding and proficiency? Do only formal assessments and tests serve as acceptable evidence? Do observations, dialogues as well as students' self-assessments can be accepted to document and validate student's achievements? What about traditional quizzes, peer reviews, peer responses, performance tasks and projects? The goal asking these questions is to determine how students will demonstrate their learning and to ensure that the assessments provide valid and reliable evidence of their achievement. This stage is critical because it ensures that the goals set in the first stage are effectively evaluated. Developing rubrics that are aligned with learning objectives and clearly communicate what constitutes different levels of performance helps students understand the expectations and provides a transparent grading process. It would be prudent to test the assessments with a small group of students to identify any issues or areas for improvement as a pilot study and then implement fully after incorporating feedback from students and colleagues to refine the assessments. This stage ensures that assessments not only measure what they are intended to but also support and enhance the overall learning experience.

3. **Plan Learning Experience and Instruction**

The third stage of backward design involves planning the learning experiences and instructional activities that will help students achieve the desired outcomes. This stage focuses on creating a coherent and engaging curriculum that supports active learning and aligns with the goals and assessments established in the previous stages. Following implementation steps are necessary for this process.
• Develop a detailed curriculum plan: This includes outline the sequence of lessons, units, and activities that will guide students toward the learning objectives. Make sure that each lesson builds on previous knowledge and skills, leading to the cumulative achievement of the desired outcomes. • Incorporate active learning strategies: Design activities that require students to actively engage with the content, such as group projects, discussions, problem-solving tasks, and hands-on experiments. Use techniques like think-pair-share, case studies, simulations, and peer teaching to foster collaboration and deeper understanding.

- Integrate formative assessments: Embed regular formative assessments within the instructional plan to monitor student progress and provide ongoing feedback. Use these assessments to inform and adjust teaching strategies, ensuring that all students are supported in their learning journey. • Differentiate instruction: Plan for diverse learning needs by incorporating differentiated instruction strategies. Provide varied resources, tasks, and supports to meet the needs of all learners, including those who may need additional challenges or remediation. Use flexible grouping, tiered assignments, and personalized learning paths to address individual differences. • Use technology and multimedia resources: Incorporate technology and multimedia resources to enhance learning and engagement. Utilize educational software, online simulations, videos, and interactive tools to complement traditional instructional methods. Ensure that these resources align with the learning objectives and are accessible to all students.
- Design for student reflection and self-assessment:

include opportunities for students to reflect on their learning and assess their own progress. This can be done through reflective journals, self-assessment checklists, and peer feedback sessions. Encourage students to set personal learning goals and develop strategies for achieving them. • Create a positive learning environment: Foster a classroom environment that encourages curiosity, risk-taking, and a growth mindset. Build a supportive community where students feel safe to express their ideas and learn from mistakes. Establish clear expectations and routines that promote respect and collaboration.

Backward design, while advantageous for aligning instruction with desired learning outcomes, presents several pitfalls. The process demands significant upfront planning, which can be time-consuming and resource-intensive, often requiring extensive professional development for educators. Its structured nature may lead to rigidity, reducing flexibility to adapt to students' varying needs and interests. Designing effective assessments that accurately measure complex learning outcomes is challenging and can inadvertently prioritize easily measurable skills over more nuanced, less tangible competencies like creativity and critical thinking. Furthermore, inexperienced educators might struggle with the complexities of backward design, and students accustomed to traditional methods may resist the structured approach. These challenges highlight the need for careful consideration and support when adopting backward design in educational settings.

## VII. DISCUSSION, OUTLOOK AND CONCLUSION

We discussed several active learning techniques specifically tailored for teaching physics courses. These methods are pivotal in fostering a more engaging and effective learning environment. Active learning, by definition, shifts the focus from passive reception of information to a more hands-on, participatory approach, which has been shown to significantly enhance students' understanding and retention of complex concepts in physics. Furthermore, the article addressed strategies to motivate students. Motivation is crucial in physics, a subject often perceived as challenging and abstract. Techniques such as connecting lessons to students' interests, using relatable examples, and fostering a growth mindset were emphasized. These approaches help students see the relevance of physics in their daily lives and encourage a positive attitude towards learning Additionally, we highlighted the importance of incorporating interactive demonstrations and simulations. These tools allow students to visualize and experiment with physical phenomena in a controlled setting, bridging the gap between abstract concepts and tangible understanding.

However, while these active learning strategies are powerful, this article also advises instructors to expand their repertoire of teaching methods. We encourage other instructors to add other approaches as well such as problem-based learning, project-based learning, experiential learning and inquiry-based learning depending on needs of student..Active learning is just one piece of the educational puzzle. Effective teaching in physics also requires thoughtful curriculum design that aligns with educational goals and standards. Instructors should continually refine their curriculum to ensure it remains relevant and comprehensive, incorporating diverse teaching methods to cater to different learning styles. We discussed briefly about curriculum design using backward design scheme, but we have not discussed about assessment and feedback. Assessments and feedback are integral to the learning process. Regular and varied assessments provide valuable insights into students' understanding and progress. These assessments should go beyond traditional exams and include formative assessments, such as quizzes, peer reviews, and reflective journals. Timely and constructive feedback is essential to guide students' learning, helping them identify areas for improvement and reinforcing their strengths. We intend to focus on assessment and feedback as a extension of this review article as next project.

Finally, we would like to note that Student Centered Active Learning Environment with Upside-down (SCALE-UP) pedagogies significantly restructure the classroom environment and pedagogies to promote highly active and interactive learning. Fostering motivation is the key to promote IPLS students' success. The development and implementation of interesting instructional activities that increase relevance and usefulness to life sciences applications, implementation of concept test as an early intervention, incorporation of diversity and inclusiveness in the classroom, and giving meaningful choices during learning process motivate IPLS students to learn. Using active learning techniques, such as think-pair-share (TPS), brainstorming, demonstrations and group work, can lead to more effective, gratifying, and memorable learning outcomes from IPLS students.

## VIII. ACKNOWLEDGMENTS

Author GS thanks Dr. Deb Hemler, Interim Dean of Science and Technology of Fairmont State University for creating conducive research and learning environment within the college.
.